\documentclass[aps,twocolumn,amsmath,amssymb,preprintnumbers,floatfix,prb,superscriptaddress,longbibliography]{revtex4-2}

\usepackage{comment}
\usepackage[version=4]{mhchem}
\usepackage[utf8]{inputenc}
\usepackage{newtxtext}
\usepackage[upint]{newtxmath}
\usepackage{microtype}
\usepackage{textcomp}
\usepackage{dsfont}
\usepackage{eucal}
\usepackage{siunitx}
\usepackage{soul}

\usepackage{enumerate}
\usepackage{amsfonts}
\usepackage{color}
\usepackage{soul}

\usepackage{todonotes}
\presetkeys%
    {todonotes}%
    {inline}{}

\usepackage{graphicx}

\usepackage[colorlinks,allcolors=blue]{hyperref}
\usepackage[capitalize]{cleveref} 
\usepackage{cleveref}

\newcommand{\vecQ}{\boldsymbol{Q}}

\newcommand{\vk}{\boldsymbol{k}}

\definecolor{DarkBlue}{rgb}{0,0,0.80}
\definecolor{DarkRed}{rgb}{0.80,0,0}
\definecolor{Purple}{rgb}{0.55,0,0.55}
\definecolor{Purple}{rgb}{0,0,0.8}

\begin{document}

\title{Mutual enhancement of altermagnetism and ferroelectricity}
\author{Chi Sun}
\affiliation{Center for Quantum Spintronics, Department of Physics, Norwegian \\ University of Science and Technology, NO-7491 Trondheim, Norway}
\author{Jacob Linder}
\affiliation{Center for Quantum Spintronics, Department of Physics, Norwegian \\ University of Science and Technology, NO-7491 Trondheim, Norway}\date{\today}
\begin{abstract}
We consider theoretically the possibility of coexisting ferroelectric and metallic altermagnetic order, which has recently
been predicted in insulating and semiconducting systems via \textit{ab initio} calculations. Solving self-consistently a mean-field Hubbard model, accounting also for the energy
cost of distorting the lattice to produce an electric polarization,
our results show that metallic altermagnetism and ferroelectricity suppress
or enhance each other depending on the doping level of the
system. Close to half-filling, the system can lower its energy by becoming
altermagnetic, but at the expense of losing the electric polarization. Away from half-filling, the coexistence of ferroelectricity
and altermagnetism is much more robust toward an increase in
the energy cost associated with the deformation of
the lattice. Therefore, our results suggest that filling fractions
corresponding to doping relatively far away from half-filling
constitute the most promising regime to look for coexistent
ferroelectricity and metallic altermagnetism with mutual enhancement. Moreover, we propose a way to electrically tune altermagnetism between nodal and nodeless phases as well as achieving coexistence of a nodal and nodeless phase for the two spin species.
\end{abstract}

\maketitle

\section{Introduction}

Ferroelectric materials are normally insulators, since a spontaneous electric polarization in a metal would just induce an electron current leading to a  redistribution of the electrons as to cancel the polarization. Therefore, ferroelectricity and metallicity at first glance seems to be two mutually incompatible properties of a material. However, a few years ago experiments demonstrated that ferroelectricity (FE) and metallicity can actually coexist in very thin layers of WTe$_2$ \cite{fei_nature_18} and MoTe$_2$ \cite{Yuan2019}.
The manner in which that FE and metallicity are able to coexist is that the metallic behavior is confined to the plane of the thin layers, whereas the polarization occurs in the out-of-plane direction. In this way, the electrons are unable to efficiently screen the polarization in the out-of-plane direction due to their confinement, whereas they can still move freely in the plane and cause the material to exhibit metallic behavior. Except for WTe$_2$ and MoTe$_2$, various metallic FE materials including L$_\text{i}$O$_\text{s}$O$_3$, PtBi$_2$, $\alpha$-In$_2$X$_3$ (X$=$S, Se, Te), and doped perovskite oxide like BaTiO$_3$ and SrTiO$_3$ have been reported \cite{shi2013,BaTiO,SrTiO,Rischau2017,TopoFE,Ding2017}. 


Recently, a newly identified magnetic phase known as altermagnetism, has drawn unprecedented interest for its unconventional symmetry and electronic structure \cite{smejkal_prx_22}, 
including its role in the fields of spintronics \cite{bai2024altermagnetism} and superconductivity \cite{liu2025altermagnetismsuperconductivityshorthistorical, cayao_jpcm_25},
and more recently multiferroics \cite{FE_switch1,FE_switch2,AFM_AM}. Altermagnets (AMs) display pronounced momentum-dependent spin splitting while retaining zero net magnetization, thereby bridging key characteristics of both ferromagnets and antiferromagnets. Meanwhile, the symmetry-protected altermagnetic band structure forbids spin-splitting along certain momentum directions constructing nodal lines. Depending on whether the nodal lines intersect the Fermi surface, AMs can be classified into two types: nodal and nodeless \cite{Jungwirth2024Altermagnets,nodeless}. While nodal AMs are extensively explored, the nodeless AMs representing a qualitatively different regime with unique electronic properties, have received much less attention.


In this work, we consider theoretically the possibility of coexisting ferroelectric and metallic altermagnetic order, which has recently been predicted in insulating and semiconducting systems via \textit{ab initio} calculations \cite{FE_switch1,FE_switch2,AFM_AM}. Solving self-consistently a mean-field Hubbard model, including the energy cost of distorting the lattice to produce an electric polarization, we find that altermagnetism and ferroelectricity will suppress or enhance each other depending on the doping level of the system. Away from half-filling, we find that altermagnetism and ferroelectricity mutually enhance each other even in a metallic state. Due to the inversion symmetry breaking in the ferroelectric state, the coexistence of altermagnetism and ferroelectricity provides a platform for investigating the interplay between non-relativistic (altermagnetism) and relativistic (Rashba SOC) effects, both of which can act as mechanisms for spin-splitting in different ways. Moreover, our results indicate a way to electrically tune the strength of the altermagnetic order and resulting spin-splitting by involving transition between nodal and nodeless phases as well as the coexistence of both phases, which is of practical interest in the context of efficient control of spintronic devices.

\section{Theory}
\begin{figure}
\includegraphics[width=0.65\columnwidth]{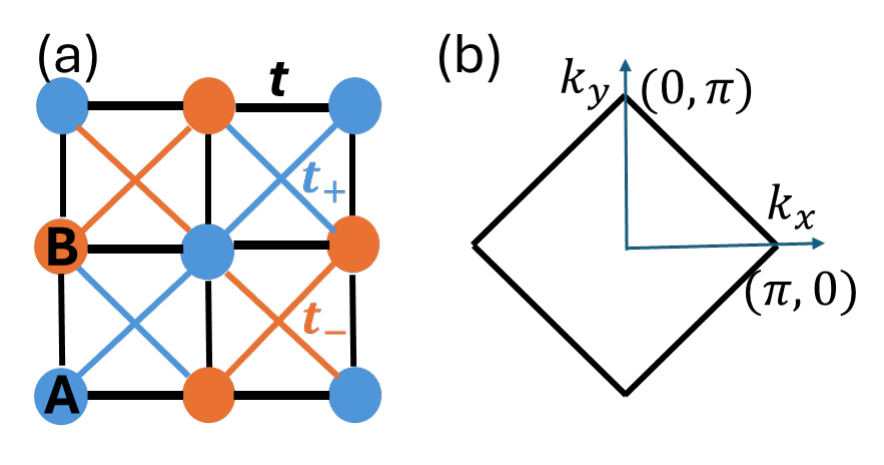}
	\caption{(a) The schematic diagram of the model Hamiltonian with sublattices A and B and (b) the associated magnetic Brillouin zone. } 
    \label{fig:MBZ}
\end{figure}

We consider a minimal model for a Rashba spin-orbit coupled ferroelectric altermagnetism in a 2D electron system based on a square lattice composed of sublattice sites A and B. Our Hamiltonian reads
\begin{align}
    H=&-\sum_{ij\sigma}t_{ij}c_{i\sigma}^\dag c_{j\sigma} -\mu\sum_{i\sigma} c_{i\sigma}^\dag c_{i\sigma}+ U\sum_i n_{i\uparrow}n_{i\downarrow}\notag\\&-\frac{i\lambda_R}{2}\sum_{ij\sigma\sigma^{'}}\hat{z}\cdot(\hat{\boldsymbol{\sigma}}\times\hat{d}_{ij})_{\sigma\sigma^{'}}c_{i\sigma}^\dag c_{j\sigma^{'}},
\end{align}
in which $c_{i\sigma}^\dag$ ($c_{i\sigma}$) is the creation (annihilation) operator at site $i$ with spin $\sigma$. The index $i$ sums over $2N=L_xL_y$ sites in $N$ unit cells. Here, $t_{ij}$ is the hopping parameter between neighboring sites, $\mu$ is the chemical potential, $U$ is the on-site Hubbard repulsion, $n_{i\sigma}=c_{i\sigma}^\dag c_{i\sigma}$, $\lambda_R$ is the Rashba coupling strength, $\hat{\boldsymbol{\sigma}}$ is the Pauli matrix vector, and $\hat{d}_{ij}$ is the nearest-neighbor vector defined as $\hat{d}_{ij}\equiv \hat{x}(\delta_{i,j-\hat{x}}-\delta_{i,j+\hat{x}})+\hat{y}(\delta_{i,j-\hat{y}}-\delta_{i,j+\hat{y}})$. Specifically, as shown in Fig. \ref{fig:MBZ}(a), we utilize uniform $t_{ij}=t$ for nearest neighbors, $t_{ij}=t_{+}$ for off-diagonal (diagonal) neighbors on the A (B) sublattice, $t_{ij}=t_{-}$ for diagonal (off-diagonal) neighbors on the A (B) sublattice, and zero elsewhere. The alternating diagonal hopping is parameterized by $t_{\pm}=t^{'}(1\pm\delta)$ with $\delta$ being the dimensionless discrepancy.

Next we apply the mean-field approximation for the Hubbard term as $n_{i\uparrow}n_{i\downarrow}\approx n_{i\uparrow}\langle n_{i\downarrow}\rangle + \langle n_{i\uparrow}\rangle n_{i\downarrow} - \langle n_{i\uparrow}\rangle \langle n_{i\downarrow}\rangle$. Use the ansatz for $\langle n_{i\sigma}\rangle$ in the spin-density-wave approach at filling $n_0$, we have \cite{Hubbard_AM1}
\begin{equation}
    \langle n_{l\lambda \sigma}\rangle=\frac{n_0}{2}+\sigma m_N e^{-i\vecQ\cdot\boldsymbol{r}_{l\lambda}}=\frac{n_0}{2}+\lambda\sigma m_N,
    \label{eq:SDW1}
\end{equation}
in which $l$ is the unit cell index and $\lambda$ is the sublattice index, which corresponds to $\sum_i=\sum_{l\lambda}$. Here we use $\lambda=+1$ ($-1$) for A (B) sublattice and $\sigma=+1$ ($-1$) for spin $\uparrow$ ($\downarrow$). $\vecQ=(\pi,\pi)$ is the magnetic ordering wave vector of the Neel state. $m_N$ is the associated Neel or
staggered magnetization, which is given by
\begin{equation}
    m_N=\frac{1}{4N}\sum_l\langle n_{lA\uparrow}\rangle-\langle n_{lA\downarrow}\rangle-\langle n_{lB\uparrow}\rangle+\langle n_{lB\uparrow}\rangle.
    \label{eq:self}
\end{equation}
A non-zero $m_N$ corresponds to a sublattice Neel ordering which gives rise to altermagnetism. Therefore, $m_N$ is treated as the altermagnetism order parameter. After Fourier transformation, we obtain
\begin{equation}
  m_N=\frac{1}{4N}\sum_{\vk}\langle n_{\vk A\uparrow}\rangle-\langle n_{\vk A\downarrow}\rangle-\langle n_{\vk B\uparrow}\rangle+\langle n_{\vk B\uparrow}\rangle, 
\end{equation}
in which $\vk$ sums over the magnetic Brillouin zone as shown in Fig. \ref{fig:MBZ}(b) .

After Fourier transformation of all terms, the mean-field Hamiltonian can be written in terms of the basis $\Psi_{\vk}^\dag=(c_{\vk A\uparrow}^\dag,c_{\vk B\uparrow}^\dag,c_{\vk A\downarrow}^\dag,c_{\vk B\downarrow}^\dag)$ as 
\begin{equation}
    H_{\text{MF}}=\sum_{\vk}\Psi^\dag_{\vk} H_{\vk}\Psi_{\vk}+E_0,
\end{equation}
in which $E_0=-2NU(\frac{n_0^2}{4}-m_N^2)$. The derivation details are given in Appendix. The $4\times4$ matrix is obtained as
\begin{equation}
    H_{\vk}=\begin{pmatrix}
\epsilon_{\vk}^{+}+\mu_{-} & \epsilon_{\vk}^0 & 0&\lambda_R^{+}\\
\epsilon_{\vk}^0 & \epsilon_{\vk}^{-}+\mu_{+}& \lambda_R^{+}&0\\0&\lambda_R^{-}&\epsilon_{\vk}^{+}+\mu_{+}&\epsilon_{\vk}^0\\\lambda_R^{-}&0&\epsilon_{\vk}^0&\epsilon_{\vk}^{-}+\mu_{-}
\end{pmatrix},
\label{eq:matrixH_R}
\end{equation}
in which $\lambda_R^{\pm}=\lambda_R(\sin{k_y}\pm i\sin{k_x})$, $\epsilon_{\vk}^0=-2t(\cos k_x+\cos k_y)$, $\epsilon_{\vk}^{\pm}=-2t_{+}\cos(k_x\pm k_y)-2t_{-}\cos(k_x \mp k_y)$, and $\mu_{\pm}=U(\frac{n_0}{2}\pm m_N)-\mu$.
Consequently, $H_{\vk}$ is diagonalized as
\begin{equation}
    H_{\vk}=\sum_{n,\vk}E_{n\vk}\gamma_{n\vk}^\dag \gamma_{n\vk}+E_0,
\end{equation}
where the new fermion operators are related to
the original ones via the relations
$
c_{\vk A\uparrow}=\sum_{n}u_{n\vk}\gamma_{n\vk}$, $\quad c_{\vk B\uparrow}=\sum_{n}v_{n\vk}\gamma_{n\vk}$,
$c_{\vk A\downarrow}=\sum_n \omega_{n\vk}\gamma_{n\vk}$, and $\quad c_{\vk B\downarrow}=\sum_n x_{n\vk}\gamma_{n\vk}
$
with $E_{n\vk}$ being the corresponding eigenvalue of the eigenvector $(u_{n\vk},v_{n\vk},\omega_{n\vk},x_{n\vk})^T$. We then proceed to self-consistently solve the altermagnetism order parameter $m_N$ and the chemical potential $\mu$ simultaneously based on the following two equations:
\begin{align}
    m_N&=\frac{1}{4N}\sum_{n\vk}(|u_{n\vk}|^2-|v_{n\vk}|^2-|\omega_{n\vk}|^2+|x_{n\vk}|^2)f(E_{n\vk}),\\
    n_0&=\frac{1}{2N}\sum_{n\vk}(|u_{n\vk}|^2+|v_{n\vk}|^2+|\omega_{n\vk}|^2+|x_{n\vk}|^2)f(E_{n\vk}),
    \label{eq:self_n}
\end{align}
in which $\langle \gamma_{n\vk}^{\dag}\gamma_{n\vk}\rangle=f(E_{n\vk})$ and $f(E_{n\vk})$ represents the Fermi distribution. Note that $\mu$ can determine the overall doping $n_0$ level of the system, here we instead consider fixed $n_0$ and solve the corresponding $\mu$ for an isolated system. 

Given the solutions of $m_N$ and $\mu$, the Helmholtz free energy per site can be calculated as
\begin{equation}
    F_0=-\frac{1}{2\beta N}\sum_{n\vk }\ln{(1+e^{-\beta E_{n\vk }})}+\frac{E_0}{2N}+\mu n_0,
    \label{eq:Helm}
\end{equation}
in which $\beta$ is the inverse temperature. Note the last term $
\mu n_0$ exists since the carrier density (filling) $n_0$ is fixed. On the other hand, the Rashba lattice polarization energy is introduced as \cite{kanasugi_prb_18}
\begin{equation}
    F_P=\frac{1}{2}\gamma P^2 + \eta P^4,
\end{equation}
where $P=C\lambda_R$ is the polarization proportional to the Rashba SOC strength. In a metallic state, the ferroelectric order develops as the emergence of a polar axis and the loss of inversion symmetry in the presence of itinerant electrons. In the presence of Rashba SOC with spontaneous inversion-symmetry breaking, the ferroelectric order is accompanied by the onset of antisymmetric SOC, i.e., $P=C\lambda_R$ is assumed.  $\gamma$ and $\eta$ are coefficients describing the elasticity of the lattice. The total free energy of the system is given by
\begin{equation}
    F_\text{tot}=F_0+F_P.
    \label{eq:tot}
\end{equation}
We then solve the lattice
polarization $P_s$ minimizing the total free energy. 

\section{Results and discussion}

\subsection{How does Rashba spin-orbit coupling affect altermagnetism?}

\begin{figure*}[ht]
 \centering
\includegraphics[width=0.7\linewidth]{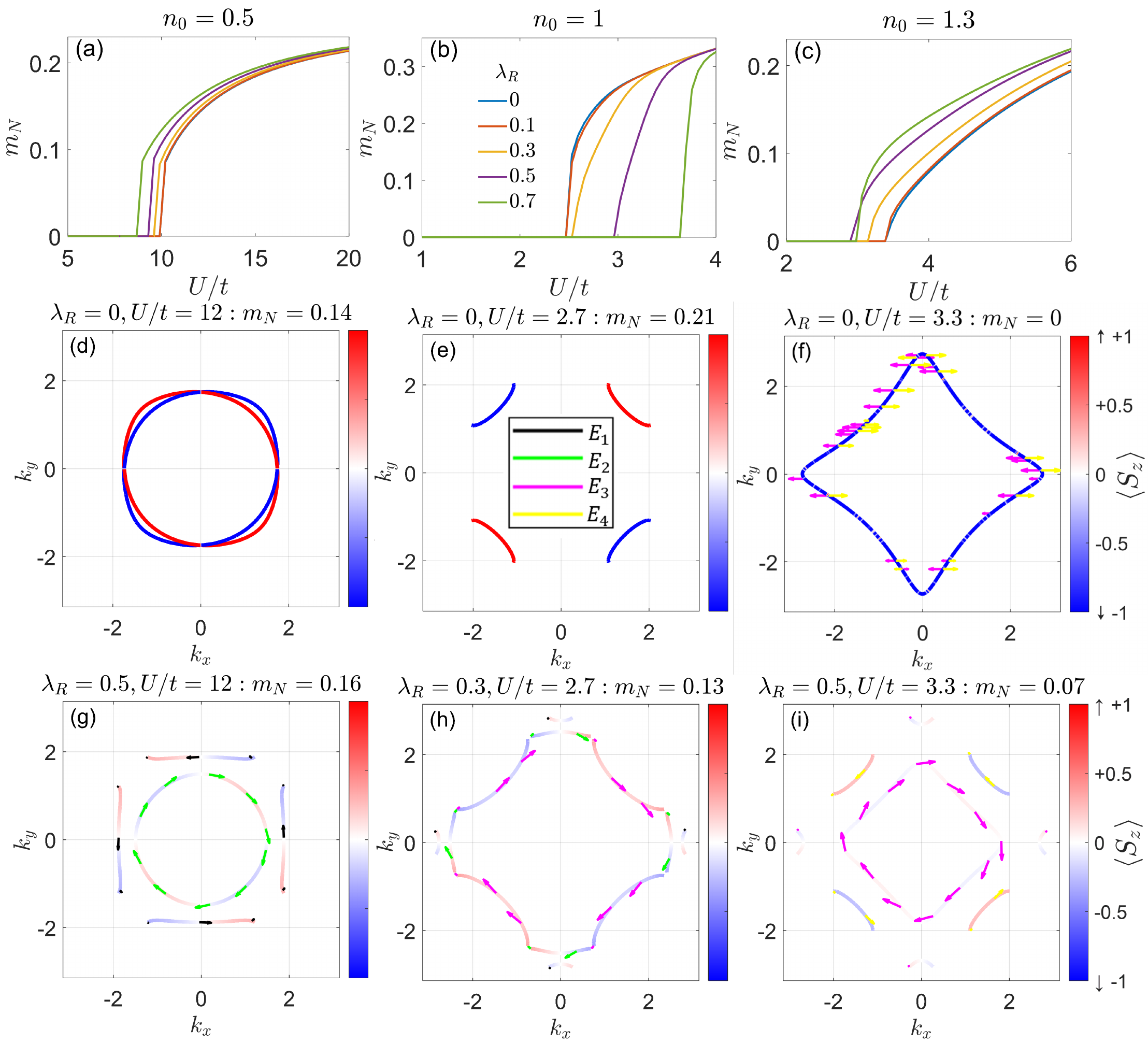}
 \caption{$m_N$ vs $U$ for different carrier density $n_0$ in the first row and the corresponding Fermi surfaces in the second (third) row in the absence (presence) of Rashba SOC $\lambda_R$. The four eigenenergies satisfy $E_1<E_2<E_3<E_4$, which can be identified by the color of the in-plane spin arrows. The parameters used are: $t=1$, $t^{'}/t=0.3$, $\delta=0.2$, and $T/t=0.1$. In panels (d) and (e), there is no in-plane component of the spin expectation value at the Fermi surface and hence the individual contributions $E_1-E_4$ cannot be discerned. Similar to (g), the inner (outer) FS in the presence of nodes corresponds to $E_2$ ($E_1$) in (d). In (e), the FS corresponds to $E_3$, the same as the magenta arrows shown in (h). }
  \label{fig:n0}
\end{figure*}

Since altermagnetism can inherently coexist with Rashba SOC in substrate-based thin films, it is essential to understand how SOC alters the magnetic behavior of altermagnets. It is known that Rashba SOC breaks inversion symmetry and gives relativistic spin-splitting, while the altermagnetism breaks
time-reversal symmetry and corresponds to non-relativistic spin-splitting effects. Their
interaction is therefore expected to result in the creation of
intricate, chiral spin structures.

In the first row of Fig. \ref{fig:n0}, we plot the self-consistently solved altermagnetism parameter $m_N$ as a function of the Hubbard repulsion $U$ in the absence and presence of Rashba SOC at different doping levels. In the absence of Rashba SOC, it is shown that a non-zero $m_N$ can be established by increasing $U$. Although not tunable in a solid state material, $U$ is experimentally tunable by Feshbach resonances in ultracold
atomic systems where lattice Hamiltonians such as ours can be designed and studied \cite{tune_U}. Similar trends has also been reported in Ref. \cite{Hubbard_AM1} at half-filling $n_0=1$. In the presence of Rashba SOC, it is found that $m_N$ decreases as $\lambda_R$ increases at a fixed $U$ at $n_0=1$, indicating the suppression of altermagnetism by Rashba SOC. Conversely, when tuning $n_0$ away from half-filling to either higher or lower region, $m_N$ can be enhanced by increasing $\lambda_R$. To unveil the underlying physics, we then look into the Fermi surfaces at different doping levels.


We start from the low doping $n_0=0.5$ (left column of Fig. \ref{fig:n0}. In the absence of SOC, the Fermi surface with two distinct ellipses [Fig. \ref{fig:n0}(d)] corresponds to a typical nodal $d$-wave AM \cite{Jungwirth2024Altermagnets}, allowing spins on each ellipse to be straightforwardly identified as solid pure red (blue) for $\langle S_z\rangle=+1$ $(-1)$ with the nodal lines appearing at $k_x=0$ and $k_y=0$. Meanwhile, the spin-split energy bands with $\langle S_z\rangle=\pm1$ almost overlap with each other, giving rise to a relatively small $m_N$. As $\lambda_R$ increases, it is seen in Fig. \ref{fig:n0}(g) that the relativistic SOC lifts the spin degeneracy that is protected at the nodes by nonrelativistic spin symmetry \cite{krempasky2024altermagnetic}, resulting in the separation of different energy iso-surfaces with the previous nodal lines fully gapped. In addition, the dominant spin expectation value $\langle S_z\rangle$ (Neel-vector-aligned expectation \cite{RKKY_SOC_AM}) is suppressed as shown by the dilution of the pure blue or red color. In each band, a characteristic alternation of spin polarization $\langle S_z\rangle$ and spin-degenerate nodes (between diluted blue and red) is shown, thus satisfying the $C_4$ symmetry and corresponding to a nodal $d$-wave AM itself. This gives a higher $m_N$ compared with the case without SOC. Similar Fermi surfaces have been shown in \cite{RKKY_SOC_AM,trama2024nonlinearanomalousedelsteinresponse,tunable_optical}, where a low energy effective altermagnetism term [i.e., $\sim m_N(k_x^2-k_y^2)\sigma_z$ or $m_Nk_xk_y\sigma_z$] is inserted as an input with the SOC term into the Hamiltonian, therefore unable to quantify the modulated-altermagnetim $m_N$ by Rashba SOC without self-consistence calculation.  We note that altermagnetism only sets in at very large values $U/t\gg 1$ for this filling fraction. For $U/t \sim 10$, it is reasonable to expect that other phases, like a Mott insulating one, becomes favorable. In what follows, we therefore increase the filling fraction $n_0$ in which case altermagnetism sets in at much lower values of $U$.

We thus consider here half-filling $n_0=1$ (middle column). As shown in Fig. \ref{fig:n0}(e) without $\lambda_R$, vanishing spin-splitting is achieved along
the $k_x=0$ and $k_y=0$ directions, resulting in symmetry-protected
altermagnetic nodal lines. Since these
nodal lines do not intersect the Fermi surface, nodeless altermagnetism is obtained, where purely out-of-plane spin polarization $\langle S_z\rangle$ with alternating signs exist. This anisotropic $\langle S_z\rangle=\pm1$ gives a large $m_N=0.21$. On the other hand, the introduction of $\lambda_R$ in Fig. \ref{fig:n0}(h) results in spin-mixed states with helical in-plane spin textures and decreases the amplitude of the out-of-plane spin polarization $\langle S_z\rangle$, same as the dilution effect observed in Fig. \ref{fig:n0}(g). In contrast, the Fermi surface is closed with alternating $\langle S_z\rangle$ signs and intersects with the nodal lines along $k_x=0$ and $k_y=0$, approaching a nodal AM with smaller $m_N$. Note here Rashba SOC introduces spin degeneracy rather than lifts it as usual, resulting in the transition from nodeless AM to nodal AM. It is shown that nodeless AM exhibits strong spin–valley polarization, resulting in completely spin-polarized electronic states within each valley \cite{nodeless}. Similarly, here we find that the transition from nodeless AM into nodal AM diminishes the altermagnetism. Unlike the transition within AM phases, the nodal to nodeless transition from AM to ferromagnet phases has also been reported by applying a magnetic field \cite{nodal_to_nodeless}.

Now we turn to the higher filling $n_0=1.3$ (right column). As shown in Fig. \ref{fig:n0}(f), no altermagnetism is achieved at $\lambda_R=0$ since the two energy bands (blue and red) are degenerate with opposite spin (both in-plane and out-of-plane). As $\lambda_R$ increases, the degeneracy between different energy bands is lifted. In each band [see Fig. \ref{fig:n0}(i)], $\langle S_z\rangle$ with alternating signs appear, which gives nonzero $m_N$. Note here the coexistence of nodal and nodeless AM is achieved: the nodal AM (carrying magenta in-plane spin arrows) possesses smaller $\langle S_z\rangle$ amplitude (more diluted color) and the nodeless AM (carrying yellow arrows) has larger $\langle S_z\rangle$ amplitude (less diluted color). 

Summarizing this section, we find that SOC strongly suppresses altermagnetism close to half-filling, whereas moving to the either of the electron- or hole-doped regime causes SOC to enhance the altermagnetic order in the system. These distinct trends are found to be associated with a transition between nodal and nodeless AMs. Note that in the widely used low energy effective altermagnetic Hamiltonian [$\sim m_N(k_x^2-k_y^2)\sigma_z$ or $m_Nk_xk_y\sigma_z$] in which the altermagnetism strength $m_N$ directly enters, strong (weak) $m_N$ (compared with the strength of the kinetic Hamiltonian term) is combined with a nodeless (nodal) altermagnetic Fermi surface \cite{strong_weak,Andreev_Chi,AM_SP}. Accordingly, in our Hubbard model where $m_N$ is solved consistently, the obtained transition from nodeless to nodal AM induced by the introduction of Rashba SOC gives a suppressed $m_N$ and vice versa.

\subsection{How does altermagnetism affect the ferroelectric polarization?}

\begin{figure}
\includegraphics[width=\columnwidth]{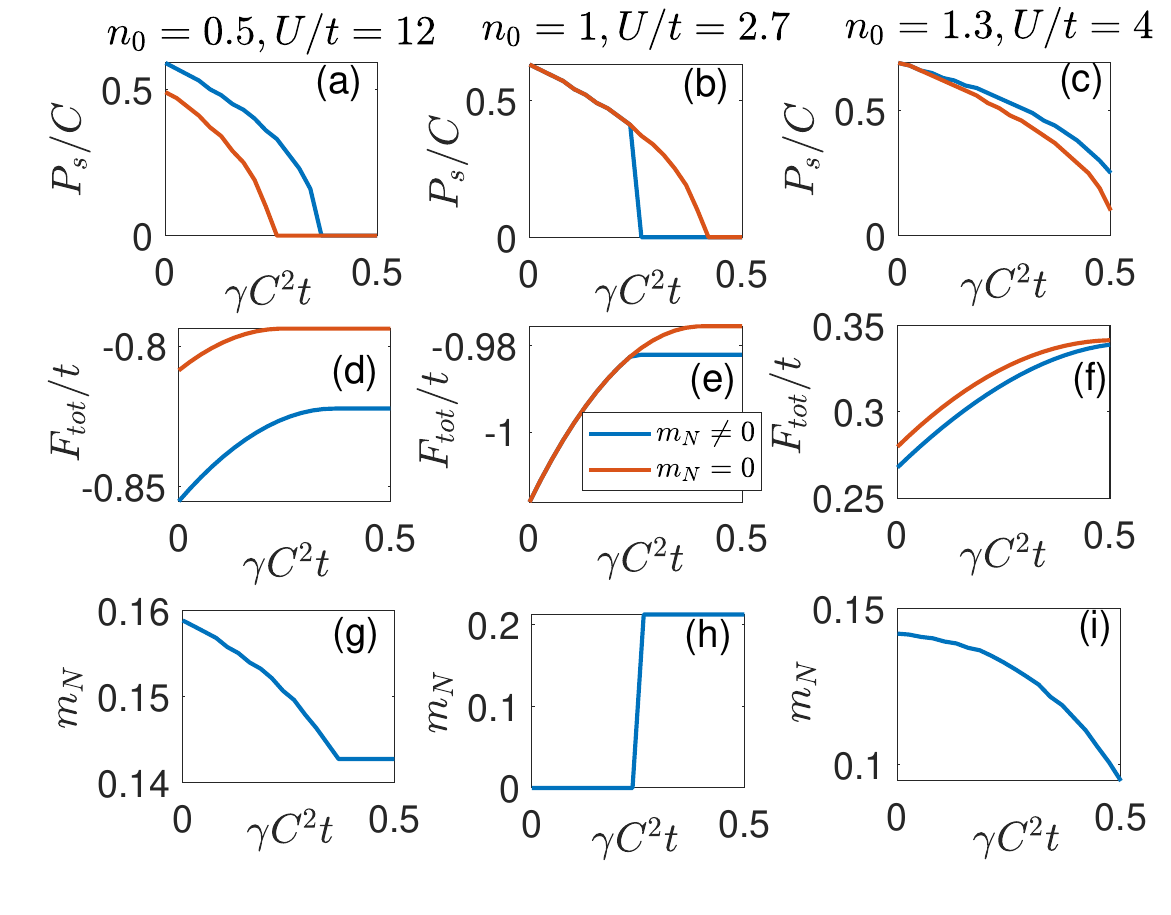}
	\caption{$P_s$ vs $\gamma$ for different carrier density $n_0$ in the first row and the corresponding free energy in the second row and $m_N$ in the third rows. The parameters used are: $t=1$, $t^{'}/t=0.3$, $\delta=0.2$, $T/t=0.1$, and $\eta C^4t^3=0.25$. } 
    \label{fig:Ps}
\end{figure}

Now we look into how the Rashba SOC-modulated altermagnetism $m_N$ affect the ferroelectricity (spontaneous electric polarization). Given the self-consistent solution of $m_N$ and $\mu$, the Helmholtz free energy is obtained by Eq. (\ref{eq:Helm}). In Fig. \ref{fig:Ps}, we include the lattice polarization contribution $F_P$ to the Helmholtz free energy to identify whether the thermodynamic ground-state is altermagnetic or not in the presence of an electric polarization $P$. We again consider the $h$-doped ($n_0=0.5$), half-filling ($n_0=1$) and $e$-doped ($n_0=1.3$) regimes. The system admits two solutions at a given value of our parameters: one with finite altermagnetism $(m_N\neq0)$ and one without altermagnetism $(m_N=0)$. In the case of $m_N=0$, only $\mu$ in Eq. (\ref{eq:self_n}) is solved consistently to calculate the free energy. To find the ground-state, we compute the free energy for both of these solutions and identify the smallest one. We emphasize that since we choose $\gamma,\eta$ to be positive, the ground-state of $F_P$ alone is $P=0$. Hence, any non-zero electric polarization arises in our model due to an energy gain from how the quasiparticles occupy the energy bands. In this way, altermagnetism, which modifies the form of the energy bands, will be shown to induce an electric polarization $P$.

The upper row in Fig. \ref{fig:Ps} shows the value $P=P_s$ of the polarization $P$ which gives the lowest free energy for both the non-altermagnetic (red curve) and altermagnetic (blue curve) solution of the self-consistency equations. Away from half-filling, the polarization is moderately higher in the altermagnetic case, showing that the ferroelectricity and altermagnetism are synergetic. At half-filling, however, the onset of  altermagnetism causes $P_s$ to drop quickly upon increasing the coefficient $\gamma$ determining the energy cost associated with deformation of the lattice.

In the middle row, we compare the free energy $F_\text{tot}$ for both the non-altermagnetic and altermagnetic solutions. Away from half-filling, the altermagnetic solution is always preferable. At half-filling, there is a crossover. At small $\gamma$, the system prefers the non-altermagnetic state, whereas increasing $\gamma$ induces a transition to an altermagnetic state. The corresponding self-consistently obtained values for $m_N$ in the ground-state of the system are shown in the lower row of Fig. \ref{fig:Ps}. As seen, $m_N=0$ at half-filling for small $\gamma$.

A key physical insight that can be taken from Fig. \ref{fig:Ps} is that close to half-filling, the system can lower its energy by becoming altermagnetic, but at the expense of losing the electric polarization. This gives a suppression of FE by altermagnetism. Similar trends have also reported in Ref. \cite{trama2024nonlinearanomalousedelsteinresponse} where the altermagnetic order suppresses the
canonical linear in-plane Rashba-Edelstein response but without inducing FE. Away from half-filling, the coexistence of ferroelectricity and altermagnetism is much more robust toward an increase in the energy cost coefficient $\gamma$ associated with the deformation of the lattice. Therefore, our results suggest that filling fractions corresponding to doping relatively far away from half-filling constitute the most promising regime to look for coexistent ferroelectricity and altermagnetism, where the spontaneous FE is enhanced by the introduction of altermagnetism.

\section{Concluding remarks}
The ferroelectric altermagnet has been predicted by using group theory and first principles \cite{FE_switch1,FE_switch2,AFM_AM}. They find that changing FE direction by applying an electric field can switch the AM polarity, but without modulating the AM strength. In our work, we focus on the interaction between altermagnetism and spontaneous ferroelectricity without an additional external electric field, and find a mutual enhancement of FE and AM. On the other hand, we arrive at a metallic state while insulating or semiconducting systems are investigated in Refs. \cite{FE_switch1,FE_switch2,AFM_AM}. Including an external electric field in our model would promote a finite value of $P$, which according to our predictions would enhance the altermagnetic order in the system away from half-filling. In this way, electrical tuning of the altermagnetic spin-splitting is achieved, even in a metallic state.
  Finally, we note that a previous work has explored the coexistence of ferroelectricity and superconductivity \cite{kanasugi_prb_18}, which enables the interesting prospect of tuning the superconducting state electrically via the out-of-plane polarization. However, no enhancement of FE is observed by the presence of superconductivity.

\acknowledgments

S. M. Selbach and J. B. Tjernshaugen are thanked for useful discussions. This work was supported by the Research
Council of Norway through Grant No. 353894 and its Centres
of Excellence funding scheme Grant No. 262633 “QuSpin.” Support from
Sigma2 - the National Infrastructure for High Performance
Computing and Data Storage in Norway, project NN9577K, is acknowledged.

\appendix
\begin{widetext}

\section{Derivation details for the mean-field Hamiltonian}

Apply the mean-field approximation for the Hubbard term:
\begin{equation}
   n_{i\uparrow}n_{i\downarrow}=n_{i\uparrow}\langle n_{i\downarrow}\rangle + \langle n_{i\uparrow}\rangle n_{i\downarrow} - \langle n_{i\uparrow}\rangle \langle n_{i\downarrow}\rangle.
\end{equation}
Use the ansatz for $\langle n_{i\sigma}\rangle$ in the spin-density-wave (SDW) approach at filling $n_0$, we write down
\begin{equation}
    \langle n_{l\lambda \sigma}\rangle=\frac{n_0}{2}+\sigma m_N e^{-i\vecQ\cdot\boldsymbol{r}_{l\lambda}},
    \label{eq:SDW1}
\end{equation}
in which $l$ is the unit cell index and $\lambda$ is the sublattice index, which gives $\sum_i=\sum_{l\lambda}$. The index $i$ sums over $2N=L_xL_y$ sites in $N$ unit cells, i.e., $l$ sums over $N$ unit cells. $\sigma=+1$ ($-1$) for $\uparrow$ ($\downarrow$).
$m_N$ is the Neel or
staggered magnetization, which is taken into account as a variational parameter. $\vecQ=(\pi,\pi)$ is the magnetic ordering wave vector of the Neel state. With $\vecQ=(\pi,\pi)$, 
\begin{equation}
  e^{-i\vecQ\cdot\boldsymbol{r}_{l\lambda}}=\lambda,\quad\sum_l e^{-i\vecQ\cdot\boldsymbol{r}_{l\lambda}}=\lambda N  
  \label{eq:SDW2}
\end{equation}
is assumed with $\lambda=+1(-1)$ for A (B) sublattice. [This comes from $e^{in\pi}=\cos(n\pi)$ gives $+1$ ($-1$) when $n$ is even (odd). Then we assume even (odd) position for A (B).] Perform the sum of $\langle n_{l\lambda \sigma}\rangle$ described by Eq. (\ref{eq:SDW1}) over $l$ and use the relation in Eq. (\ref{eq:SDW2}), we obtain
\begin{equation}
    \langle n_{l\lambda\sigma}\rangle=\frac{n_0}{2}+\lambda\sigma m_N, \quad\sum_l\langle n_{l\lambda\sigma}\rangle=N(\frac{n_0}{2}+\lambda\sigma m_N),
    \label{eq:SDW_sum}
\end{equation}
which corresponds to $\sum_l\langle n_{lA\uparrow}\rangle=N(\frac{n_0}{2}+m_N)$, $\sum_l\langle n_{lA\downarrow}\rangle=N(\frac{n_0}{2}-m_N)$,  $\sum_l\langle n_{lB\uparrow}\rangle=N(\frac{n_0}{2}-m_N)$ and $\sum_l\langle n_{lB\downarrow}\rangle=N(\frac{n_0}{2}+m_N)$. Based on the above, we obtain
\begin{align}
    m_N&=\frac{1}{4N}\sum_l\langle n_{lA\uparrow}\rangle-\langle n_{lA\downarrow}\rangle-\langle n_{lB\uparrow}\rangle+\langle n_{lB\uparrow}\rangle,
    \label{eq:self}\\
    n_0&=\frac{1}{2N}\sum_l \langle n_{lA\uparrow}\rangle+\langle n_{lA\downarrow}\rangle+\langle n_{lB\uparrow}\rangle+\langle n_{lB\uparrow}\rangle
\end{align}
which gives the self-consistent equations for the order parameter $m_N$ and chemical potential $\mu$. After Fourier transformation, they become
\begin{align}
  m_N=\frac{1}{4N}\sum_{\vk}\langle n_{\vk A\uparrow}\rangle-\langle n_{\vk A\downarrow}\rangle-\langle n_{\vk B\uparrow}\rangle+\langle n_{\vk B\uparrow}\rangle,\\
  n_0=\frac{1}{2N}\sum_{\vk}\langle n_{\vk A\uparrow}\rangle+\langle n_{\vk A\downarrow}\rangle+\langle n_{\vk B\uparrow}\rangle+\langle n_{\vk B\uparrow}\rangle,
\end{align}
in which $\vk$ sums over the magnetic Brillouin zone. Insert Eq. (\ref{eq:SDW_sum}) and perform Fourier transformation $c_{l\lambda\sigma}^\dag=\frac{1}{\sqrt{N}}\sum_{\vk}c_{\vk\lambda\sigma}^\dag e^{-i\vk\cdot\boldsymbol{r}_{l\lambda}}$, we arrive at
\begin{align}
    \sum_{l\lambda}n_{l\lambda\uparrow}\langle n_{l\lambda\downarrow}\rangle&=\sum_{\vk\lambda}(\frac{n_0}{2}-\lambda m_N)c_{\vk\lambda\uparrow}^\dag c_{\vk\lambda\uparrow}=\sum_{\vk} (\frac{n_0}{2}- m_N)c_{\vk A\uparrow}^\dag c_{\vk A\uparrow}+(\frac{n_0}{2}+ m_N)c_{\vk B\uparrow}^\dag c_{\vk B\uparrow},\\
    \sum_{l\lambda}\langle n_{l\lambda\uparrow}\rangle n_{l\lambda\downarrow}&=\sum_{\vk\lambda}(\frac{n_0}{2}+\lambda m_N)c_{\vk\lambda\downarrow}^\dag c_{\vk\lambda\downarrow}=\sum_{\vk} (\frac{n_0}{2}+ m_N)c_{\vk A\downarrow}^\dag c_{\vk A\downarrow}+(\frac{n_0}{2}- m_N)c_{\vk B\downarrow}^\dag c_{\vk B\downarrow},\\
    \sum_{l\lambda}\langle n_{l\lambda\uparrow}\rangle \langle n_{l\lambda\downarrow}\rangle&=\sum_{l\lambda}(\frac{n_0}{2}+\lambda m_N)(\frac{n_0}{2}-\lambda m_N) =2N(\frac{n_0^2}{4}-m_N^2).
\end{align}

We then turn to the spacial Hamiltonian of the Rashba term,
\begin{equation}
    H_R=-\frac{i\lambda_R}{2}\sum_{ij\sigma\sigma^{'}}\hat{z}\cdot(\hat{\boldsymbol{\sigma}}\times\hat{d}_{ij})_{\sigma\sigma^{'}}c_{i\sigma}^\dag c_{j\sigma^{'}},
    \label{eq:spacial_Rashba}
\end{equation}
where $\lambda_R$ is the Rashba strength and $\hat{\boldsymbol{\sigma}}$ is the Pauli matrix vector. Here we consider the inversion symmetry breaking along $\hat{z}$ and the 2D square lattice in the $x$-$y$ plane. $\hat{d}_{ij}$ is the nearest-neighbor vector defined as
\begin{equation}
    \hat{d}_{ij}\equiv \hat{x}(\delta_{i,j-\hat{x}}-\delta_{i,j+\hat{x}})+\hat{y}(\delta_{i,j-\hat{y}}-\delta_{i,j+\hat{y}}).
\end{equation}
For the Neel state considered here, we have $i\in A$ with $j\in B$ and vice versa for the nearest neighbor. The sum over nearest neighbors can be written as
\begin{equation}
    \sum_{ij}=\sum_{i,j=i+\hat{x}}+\sum_{i,j=i-\hat{x}}+\sum_{i,j=i+\hat{y}}+\sum_{i,j=i-\hat{y}},   
\end{equation}
which corresponds to $\hat{d}_{ij}=\hat{x}$, $-\hat{x}$, $\hat{y}$ and $-\hat{y}$ for the four terms, respectively. The four terms after Fourier transformation are calculated as 
\begin{align}
    \sum\limits_{\substack{i,j=i+\hat{x} \\ \sigma\sigma^{'}}}\hat{z}\cdot(\hat{\boldsymbol{\sigma}}\times\hat{x})_{\sigma\sigma^{'}}c_{i\sigma}^\dag c_{j\sigma^{'}}&=\sum_{\vk\sigma\sigma^{'}}(-\sigma_y)_{\sigma\sigma^{'}}e^{ik_x}(c_{\vk A\sigma}^\dag c_{\vk B\sigma^{'}}+c_{\vk B\sigma}^\dag c_{\vk A\sigma^{'}}),\\
    \sum\limits_{\substack{i,j=i-\hat{x} \\ \sigma\sigma^{'}}}\hat{z}\cdot[\hat{\boldsymbol{\sigma}}\times(-\hat{x})]_{\sigma\sigma^{'}}c_{i\sigma}^\dag c_{j\sigma^{'}}&=\sum_{\vk\sigma\sigma^{'}}(\sigma_y)_{\sigma\sigma^{'}}e^{-ik_x}(c_{\vk A\sigma}^\dag c_{\vk B\sigma^{'}}+c_{\vk B\sigma}^\dag c_{\vk A\sigma^{'}}),\\
\sum\limits_{\substack{i,j=i+\hat{y} \\ \sigma\sigma^{'}}}\hat{z}\cdot(\hat{\boldsymbol{\sigma}}\times\hat{y})_{\sigma\sigma^{'}}c_{i\sigma}^\dag c_{j\sigma^{'}}&=\sum_{\vk\sigma\sigma^{'}}(\sigma_x)_{\sigma\sigma^{'}}e^{ik_y}(c_{\vk A\sigma}^\dag c_{\vk B\sigma^{'}}+c_{\vk B\sigma}^\dag c_{\vk A\sigma^{'}}),\\
\sum\limits_{\substack{i,j=i-\hat{y} \\ \sigma\sigma^{'}}}\hat{z}\cdot[\hat{\boldsymbol{\sigma}}\times(-\hat{y})]_{\sigma\sigma^{'}}c_{i\sigma}^\dag c_{j\sigma^{'}}&=\sum_{\vk\sigma\sigma^{'}}(-\sigma_x)_{\sigma\sigma^{'}}e^{-ik_y}(c_{\vk A\sigma}^\dag c_{\vk B\sigma^{'}}+c_{\vk B\sigma}^\dag c_{\vk A\sigma^{'}}).
\end{align}
Sum over the above four equations and then times the prefactor $-\frac{i\lambda_R}{2}$ as shown in Eq. (\ref{eq:spacial_Rashba}), we arrive at
\begin{align}
   H_R&=\sum_{\vk\sigma\sigma^{'}}\lambda_R[(\sigma_{x})_{\sigma\sigma^{'}}\sin{k_y}-(\sigma_y)_{\sigma\sigma^{'}}\sin{k_x}](c_{\vk A\sigma}^\dag c_{\vk B\sigma^{'}}+c_{\vk B\sigma}^\dag c_{\vk A\sigma^{'}}) \notag\\&=\sum_{\vk\lambda\sigma\sigma^{'}}\lambda_R[(\sigma_{x})_{\sigma\sigma^{'}}\sin{k_y}-(\sigma_y)_{\sigma\sigma^{'}}\sin{k_x}]c_{\vk\lambda\sigma}^\dag c_{\vk,-\lambda,\sigma^{'}}.
\end{align}

After Fourier transformation of the hopping and chemical potential terms, the mean-field Hamiltonian can be written in terms of the basis $\Psi_{\vk}^\dag=(c_{\vk A\uparrow}^\dag,c_{\vk B\uparrow}^\dag,c_{\vk A\downarrow}^\dag,c_{\vk B\downarrow}^\dag)$ as $H_{\text{MF}}=\sum_{\vk}\Psi^\dag_{\vk} H_{\vk}\Psi_{\vk}+E_0$ with the expression details given in the main text.

\end{widetext}

\bibliography{bib}


\end{document}